\documentclass[aps,twocolumn]{revtex4-1}

\usepackage{amsmath,amssymb,bm}
\usepackage{color,graphicx,subfigure}
\usepackage{graphpap}
\usepackage{rotating}

\newcommand{\bB}{\mathbf{B}}

\newcommand{\bBe}{\mathbf{B}_{\mathrm{e}}}
\newcommand{\bBeC}{\bBe^{\!\blacktriangledown}}
\newcommand{\bBeI}{\mathbf{B}^{(1)}_{\mathrm{e}}}
\newcommand{\bbD}{\mathbb{D}}

\newcommand{\bbM}{\mathbb{M}}

\newcommand{\bD}{\mathbf{D}}

\newcommand{\bF}{\mathbf{F}}
\newcommand{\bFe}{\mathbf{F}_{\!\mathsf{e}}}
\newcommand{\bFeI}{\mathbf{F}^{(1)}_{\!\mathrm{e}}}
\newcommand{\bFeO}{\mathbf{F}^{(0)}_{\!\mathrm{e}}}
\newcommand{\bFer}{{}_{\eps}\mathbf{F}_{\!\mathrm{e}}}
\newcommand{\bFeres}{\mathbf{F}^{(+)}_{\!\mathrm{e}}}
\newcommand{\bFg}{\mathbf{G}}

\newcommand{\bH}{\mathbf{H}}
\newcommand{\bI}{\mathbf{I}}

\newcommand{\bPsi}{\bm{\Psi}}
\newcommand{\bPsiC}{\bPsi^{\!\blacktriangledown}}

\newcommand{\bL}{\mathbf{L}}
\newcommand{\bM}{\mathbf{M}}
\newcommand{\bn}{\mathbf{n}}
\newcommand{\bno}{\mathbf{\mathring{n}}}

\newcommand{\bT}{\mathbf{T}}
\newcommand{\bW}{\mathbf{W}}

\newcommand{\bv}{\mathbf{v}}

\newcommand{\eps}{\varepsilon}

\newcommand{\siani}{\Upsilon_{\!\mathsf{shr}}}
\newcommand{\bTani}{\bT_{\mathsf{shr}}}

\newcommand{\Ldot}[1]{\overset{\bm{.}}{#1}}

\DeclareMathOperator{\tp}{\otimes}

\DeclareMathOperator{\sym}{sym}
\DeclareMathOperator{\skw}{skw}
\DeclareMathOperator{\tr}{tr}

\DeclareMathOperator{\dev}{dev}

\newcommand {\bydef}{\,\raise.07485ex\hbox{:}\kern-.025em\hbox{=}\,}
\newcommand {\defby}{\,=\kern-.345em\hbox{\raise.07485ex\hbox{:}}\,}

\newcommand{\bna}{\boldsymbol{\nabla}}
\begin{document}

\title{Liquid relaxation: A new Parodi-like relation for nematic liquid crystals}
\author{Paolo Biscari}
\affiliation{Dipartimento di Fisica, Politecnico di Milano, Piazza Leonardo da Vinci 32, 20133 Milano, Italy}
\author{Antonio DiCarlo}
\affiliation{CECAM--IT--SIMUL Node c/o Universit\`a Roma Tre, Via Madonna dei Monti 40, 00184 Roma, Italy}
\author{Stefano S.\ Turzi}
\affiliation{\mbox{Dipartimento di Matematica, Politecnico di Milano, Piazza Leonardo da Vinci 32, 20133 Milano, Italy}}

\date{\today}

\begin{abstract}
We put forward a hydrodynamic theory of nematic liquid crystals that includes both anisotropic elasticity and dynamic relaxation. Liquid remodeling is encompassed through a continuous update of the shear-stress free configuration. The low-frequency limit of the dynamical theory reproduces the classical Ericksen-Leslie theory, but it predicts \emph{two} independent identities between the six Leslie viscosity coefficients. One replicates
Parodi's relation, while the other---which involves five Leslie viscosities in a nonlinear way---is new. We discuss its significance, and we test its validity against evidence from physical experiments, independent theoretical predictions, and molecular-dynamics simulations.
\end{abstract}

\maketitle

\section{Introduction}
Liquids are unable to sustain any nonzero \emph{stationary} shear stress. In ordinary conditions---i.e., at small enough strain rates---shear relaxation occurs exponentially fast, producing a viscoelastic analog of the dielectric Debye relaxation. However, at the crossover between the characteristic shearing time and the liquid relaxation time \cite{47fre}, distinctive solidlike features become increasingly manifest \cite{85hido07tra12bflrt}. Molecular rearrangements are dramatically slowed down in confined ultrathin liquid films (three to ten molecular dimensions thick), whose relaxation times may be as large as tens to hundreds of milliseconds \cite{91hcg}, making the crossover more experimentally accessible. But clear fingerprints of a smooth transition from liquidlike to solidlike response manifest also in the acoustic properties of nematic liquid crystals (NLCs) in the MHz--GHz frequency range \cite{95gr&95kr&99ruo&12kkkk}. \par
In this paper we show how a fairly general continuum theory of liquids may be established by allowing the \emph{effective} shear strain---i.e., the shear strain from an evolving \emph{relaxed configuration}---to enter the strain energy functional. A dissipation principle governs the evolution of such a configuration, and it takes into account the macroscopic effects of microscopic rearrangements \cite{02diq04rasi}. In a previous work \cite{14bdt15tur}, we constructed such a theory for (slightly) compressible NLCs and applied it, with fair success, to explain quantitatively the anisotropy of sound velocity \cite{72muls} and sound attenuation \cite{70lordlab} in $N$-(4-methoxybenzylidene)-4-butylaniline (MBBA) over the range 2--14 MHz. The theory of nematic relaxation put forth in \cite{14bdt15tur} is characterized by: (i) a neo-Hookean contribution to the strain energy where the effective shear strain enters weighted by an anisotropic shape tensor, and (ii) an isotropic gradient flow dynamics for the relaxed 
configuration, parameterized by a single viscosity modulus. Here we keep (i) as is, but we revise and extend (ii) by taking a gradient flow with respect to an anisotropic metric possessing the minimum symmetry compatible with the liquid crystal. This theory covers the whole range from low-frequency hydrodynamics to solidlike high-frequency regimes \cite{95gr&95kr&99ruo&12kkkk}. In particular, the low-frequency predictions reproduce the well-known Ericksen-Leslie \cite{60eri,68lesl} dynamical theory, but they deliver in addition a new Parodi-like relation between viscosity coefficients. Along with the original Parodi relation \cite{70paro} ---which we also retrieve---this result lowers to \emph{four} the number of independent viscosities for a nematic liquid crystal. Both conditions involve only (some of) the six original Leslie coefficients, not the extra three viscosities entering the extension of the Ericksen-Leslie theory to compressible NLCs \cite{75mon&}. Accordingly, only the theory for incompressible 
NLCs will be presented here, and its predictions tested against experimental data, earlier theoretical predictions, and results from molecular-dynamics (MD) simulations. The discussion of the compressible case is left to a future paper. \par
\section{Relaxational dynamics}
We briefly sketch the theory that provides the equations of motion for a NLC, under the combined effect of anisotropic elasticity and anisotropic relaxation. Let $\bF$ be the deformation gradient from an arbitrarily selected reference configuration of the NLC body. To account for relaxation, we factorize $\bF$ into a \emph{relaxing deformation} $\bFg$ and an \emph{effective deformation} $\bFe$:
\begin{equation} \label{eq:split}
\bF\!=\!\bFe\bFg\,,
\end{equation}
with $\bFg$ identifying the relaxed equilibrium configuration, and the effective deformation $\bFe$ measuring the deviation from equilibrium of the current deformation. Consequently, only the effective deformation enters the strain energy. Since $\bFe\!=\!\bF\bFg^{-1}$ maps from the relaxed to the current configuration, the strain energy is properly defined, being independent of the arbitrarily chosen reference. For an incompressible NLC, all factors in \eqref{eq:split} are isochoric, i.e., have unit determinant. \par
In order to account for anisotropic elasticity, we augment the classical Oseen-Frank \cite{O-F} free-energy density (per unit volume) with the anisotropic potential
\begin{equation} \label{eq:sheared}
\siani=\tfrac{1}{2}\mu \tr(\bPsi^{-1}\bBe - \bI),
\end{equation}
where $\mu$ is the shear modulus. The strain energy \eqref{eq:sheared} simply measures the deviation of the \emph{effective strain} $\bBe\bydef\bFe\bFe^\top$ from the \emph{energetic shape tensor}
\begin{equation} \label{eq:shape}
\bPsi\bydef a^2 (\bn \tp \bn) + a^{-1}(\bI - \bn \tp \bn),
\end{equation}
parameterized by the \emph{aspect ratio} $a\!>\!0$, whose deviation from~$1$ gauges the degree and the type (prolate or oblate) of elastic anisotropy with respect to the nematic director $\bn$ \cite{ratio}. The shape tensor $\bPsi$ is symmetric, positive definite, and with unit determinant. The potential $\siani$ adds the following contribution to the stress tensor \cite{14bdt15tur,dev}:
\begin{equation} \label{eq:stress}
\bTani\!=\!\mu\dev(\bPsi^{-1}\bBe).
\end{equation}
Clearly, $\bTani$ vanishes if and only if $\bBe\!=\!\bPsi$, where $\siani$ attains its unique minimum. \par
We now proceed to derive an evolution equation for the relaxing deformation $\bFg$. Since $\bBe\!=\bF\bH\bF^\top\!$, with the \emph{inverse relaxing strain} defined as $\bH\bydef(\bFg^\top\bFg)^{-1}$, the strain energy density \eqref{eq:sheared} depends on the relaxing deformation $\bFg$ only through $\bH$. Any relaxation dynamics necessarily obeys a dissipation inequality, ensuring a nonnegative entropy production. In this case, such an inequality reads
\begin{equation} \label{eq:ClDuin}
\big(\bF^{\!\top}\bPsi^{-1}\bF\big)\!\cdot\!\dot{\bH}\leq0
\end{equation}
(see \cite{14bdt15tur}). In terms of the \emph{co-deformational derivative}
\begin{equation}
\bBeC\bydef\dot{\bB}_{\mathsf{e}}-(\bna\bv)\,\bBe-\bBe(\bna\bv)^{\!\top}\!=\,\bF\Ldot{\bH}\bF^\top\!,
\end{equation}
where $\bv$ is the translational velocity field and $\bna\bv\!=\!\dot{\bF}\bF^{-1}$ is its spatial gradient, inequality \eqref{eq:ClDuin} takes the form
\begin{equation} \label{eq:ClDuStrin}
\bPsi^{-1}\!\cdot\bBeC\!<0\quad\textrm{unless}\quad\bBeC\!=0\,,
\end{equation}
made stricter by the presumption that relaxation does dissipate. The simplest way to satisfy it is to assume that there is an invertible \emph{dissipation tensor} $\bbD$ whose symmetric part is positive definite, such that
\begin{equation} \label{eq:Diss}
\bbD\,\bBeC=-\,\bPsi^{-1}+\lambda\,\bBe^{-1},
\end{equation}
where the Lagrange multiplier $\lambda$ enforces the condition that the relaxation process be isochoric. After introducing the \emph{mobility tensor} $\,\bbM\bydef\bbD^{-1}$, this yields the gradient-flow equation \cite{IP}
\begin{equation} \label{eq:GradFlow}
\bBeC=-\,\bbM\!\left(\!\bPsi^{-1}-
\frac{(\bbM\,\bPsi^{-1})\!\cdot\!\bBe^{-1}}{(\bbM\,\bBe^{-1})\!\cdot\!\bBe^{-1}}\,\bBe^{-1}\!\right)\!.
\end{equation}
Note that, contrary to $\bH$, the effective strain $\bBe$ is independent of the arbitrarily chosen reference. Hence, the relaxation dynamics \eqref{eq:GradFlow} is properly defined. \par
The most general dissipation tensor $\bbD$ sharing the symmetry of the shape tensor \eqref{eq:shape} may be parameterized by six scalar coefficients $\tau_1,\dots,\tau_6$, as \cite{87povi}
\begin{equation} \label{eq:Ctrasv}
\begin{aligned}
\bbD\,\bL&=\tau_1\bL+\tau_2(\tr\bL)\,\bI
+\tau_3(\bPsi\bL+\bL\bPsi)
\\
&\,+\tau_4\big((\tr\bL)\bPsi+(\bPsi\!\cdot\!\bL)\,\bI\,\big)
+\tau_5\bPsi\bL\bPsi
\\
&\,+\tau_6\big((\tr\bL)\bPsi-(\bPsi\!\cdot\!\bL)\,\bI\,\big),\quad \text{for all }\bL\in\text{Sym}.
\end{aligned}
\end{equation}
It depends on the aspect ratio $a$ via $\bPsi$, and possibly also via the coefficients $\tau_1,\dots,\tau_6$. Generically, $\bbD$ has two double eigenvalues:
\begin{equation} \label{eq:T&T}
\begin{aligned}
\tau_{\perp}&=\tau_1+(a^2\!+a^{-1})\tau_3+a\,\tau_5>0\,,
\\
\tau_{\|}&=\tau_1+2\,a^{-1}\,\tau_3+a^{-2}\,\tau_5>0\,,
\end{aligned}
\end{equation}
associated respectively with the shearing modes in the plane normal to $\bn$ and the shearing modes that tilt the nematic director. Their inverses measure how fast these modes relax. On the (complementary) invariant subspace spanned by the orthonormal pair $\big((\sqrt{3}/2)\dev(\bn\!\otimes\!\bn)$, $(1/\sqrt{3}\,)\bI\big)$, $\bbD$ acts as follows
\begin{align*}
[\,\bbD\,] =
\begin{pmatrix}
\tau_{11}&\tau_{12}\\
\tau_{21}&\tau_{22}
\end{pmatrix},
\end{align*}
where
\begin{align*}
&\tau_{11}=\tau_1
+\frac{4\,a^3+2}{3\,a}\,\tau_3
+\frac{2\,a^6+1}{3\,a^2}\,\tau_5\,,
\\
&\tau_{22}=\tau_1
+3\,\tau_2
+2\,\frac{a^3+2}{3\,a}\,\tau_3
+2\,\frac{a^3+2}{a}\,\tau_4
+\frac{a^6+2}{3\,a^2}\,\tau_5\,,
\\[.25ex]
&\tfrac12(\tau_{12}+\tau_{21})=
\sqrt{2}\,\frac{a^3-1}{3a^2}\big(2\,a\,\tau_2+3\,a\,\tau_4+(1+a^3)\,\tau_5\big)\,,
\\[.5ex]
&\tfrac12(\tau_{12}-\tau_{21})=\sqrt{2}\,\frac{a^3-1}{a}\,\tau_6\,.
\end{align*}
with
\begin{equation} \label{eq:Tij}
\tau_{11}\tau_{22}-\tfrac14\big(\tau_{12}+\tau_{21}\big)^2>0\,.
\end{equation}
Under positivity conditions \eqref{eq:T&T} and \eqref{eq:Tij}, $\bbD$ is invertible whatever the value of $\tau_6$ is. For $a\!=\!1$ (a condition identifying an isotropic liquid), $\tau_{\perp}$ and $\tau_{\|}$ collapse into $\tau_{11}\!=\tau_1\!+2\,\tau_3+\tau_5\!>0\,$, $\tau_{12}\!=\tau_{21}\!=0\,$, and \eqref{eq:Ctrasv} reduces~to
\begin{equation*}
\bbD\,\bL=\tau_{11}\dev\bL+\tau_{22}\,\tfrac13(\tr\bL)\bI\,,
\end{equation*}
with $\tau_{22}=\tau_{11}\!+3\,(\tau_2+2\,\tau_4)>0\,$. \par\smallskip
\section{Low-frequency limit: a new Parodi relation}
Now that we have characterized both the elastic and the relaxational material properties, we set up a perturbative procedure fit to study the slow motions where the system is expected to comply with the Ericksen-Leslie hydrodynamics. The evolution equation \eqref{eq:GradFlow} has only one stationary solution: $\bBe\!=\!\bPsi$, which is globally attractive. Therefore, if the deformation process is slow enough (on the time scale set by the largest relaxation time characterizing $\bbD$), the ensuing viscous response is well described by linearizing the right side of \eqref{eq:GradFlow} about $\bPsi$ and assuming the deformation gradient to be \emph{retarded} in the sense of \cite{Col&Noll}:
\begin{equation} \label{retard}
\bFer(t)\bydef\bFeO(\eps\,t)+\eps\,\bFeI(\eps\,t)+o(\eps)\bFeres(\eps\,t)\,,
\end{equation}
implying that ${}_{\eps}\dot{\bF}_{\!\mathsf{e}}\!=\eps\,\dot{\mathbf{F}}^{(0)}_{\!\mathrm{e}}\!+o(\eps)$ and ${}_{\eps}\bBeC=\eps\,\bPsiC\!+o(\eps)$. Under these assumptions, equation \eqref{eq:GradFlow}, trivially satisfied at $O(1)$, at $O(\eps)$ leads to
\begin{equation} \label{eq:LinGradFlow}
\dev\!\big(\bPsi^{-1}\bBeI\big)=-\dev\!\big((\bbD\,\bPsiC)\bPsi\big),
\end{equation}
which, substituted into \eqref{eq:stress}, yields
\begin{equation} \label{eq:linstress}
\bTani^{(1)}\!=\!-\mu\dev\!\big((\bbD\,\bPsiC)\bPsi\big).
\end{equation}
The co-deformational derivative of the shape tensor \eqref{eq:shape} reads
\begin{align*}
\bPsiC&\bydef\dot{\bPsi}-(\bna\bv)\,\bPsi-\bPsi(\bna\bv)^{\!\top}
\\
&=2\!\left(a^2\!-a^{-1}\right)\sym\!\big(\bno\!\otimes\!\bn-(\bD\bn)\!\otimes\!\bn\big)-2\,a^{-1}\bD\,,
\end{align*}
where $\bno\bydef\dot{\bn}-\bW\bn$ is the co-rotational derivative of the nematic director, $\bD\bydef\sym(\bna\bv)$ the (traceless) stretching, and $\bW\bydef\skw(\bna\bv)$ is the spin. \par
We are now in a position to compare our result \eqref{eq:linstress} with the most general linear viscous stress compatible with the nematic structure, as posited in \cite{68lesl}, namely
\begin{equation}\label{eq:TLeslie}
\begin{aligned}
\alpha_1\bn\!\cdot\!(\bD\bn)\bn\!\tp\!\bn+\alpha_2\bno\!\tp\!\bn+\alpha_3\bn\!\tp\!\bno
\\
\,+\,\alpha_4\bD+\alpha_5(\bD\bn)\!\tp\!\bn+\alpha_6\bn\!\tp(\bD\bn)\,,
\end{aligned}
\end{equation}
whose traceless component matches \eqref{eq:linstress} provided that the six Leslie viscosities are identified as follows:
\begin{align}
\alpha_1&=\mu\,a^{-3}\big[2\,(a^3-1)^2\,\big(a^7\tau_1+a\,\tau_2+2\,(a^3+1)\,\tau_4\big) \notag\\
&+a\,(4\,a^9-a^6-2\,a^3-1)\,\tau_{\perp}\!-2\,a\,(a^{12}-1)\,\tau_{\|}\big], \notag\\
\alpha_2&=-\mu\,(a^3\!-1)\,a\,\tau_{\perp}, \notag\\[.5ex]
\alpha_3&=-\mu\,(a^3\!-1)\,a^{-2}\,\tau_{\perp}, \label{eq:alphas}\\[.5ex]
\alpha_4&=2\,\mu\,a^{-2}\,\tau_{\|}, \notag\\[.5ex]
\alpha_5&=\mu\big((a^3\!+1)\,a\,\tau_{\perp}-2\,a^{-2}\,\tau_{\|}\big), \notag\\[.5ex]
\alpha_6&=\mu\big((a^3\!+1)\,a^{-2}\,\tau_{\perp}-2\,a^{-2}\,\tau_{\|}\big).\notag
\end{align}
These viscosity coefficients satisfy identically the well-known Parodi relation \cite{70paro}
\begin{equation}\label{eq:P}
\alpha_2+\alpha_3=\alpha_6-\alpha_5\,.
\end{equation}
This should be expected, since $\tau_6$, the only coefficient breaking the symmetry of $\bbD$, does not enter equalities \eqref{eq:alphas} \cite{tau6}. A far less obvious result is the new nonlinear relation involving all Leslie viscosities but $\alpha_1$:
\begin{equation}\label{eq:P-like}
\frac{\alpha_2}{\alpha_3}=\frac{\alpha_4+\alpha_5}{\alpha_4+\alpha_6}\,,
\end{equation}
and the fact that the cubic root of the two ratios equated in \eqref{eq:P-like} equals the aspect ratio $a$:
\begin{equation}\label{eq:P-a}
\frac{\alpha_2}{\alpha_3}=\frac{\alpha_4+\alpha_5}{\alpha_4+\alpha_6}=a^3.
\end{equation}
For $a\!=\!1$ (implying $\bPsi\!=\!\bI$ and an \emph{isotropic} free-energy density), all $\alpha$'s vanish but $\alpha_4\!=2\,\mu\,\tau_{11}$. The \mbox{ratio $\alpha_2/\alpha_3$} is hence undefined. However, \eqref{eq:P-like} and \eqref{eq:P-a} still hold by continuity. Parodi's relation \eqref{eq:P}, stemming from a general thermodynamic argument, is so well established that it is simply taken for granted by experimentalists who identify all of the six Leslie coefficients in the absence of data from normal stress measurements \cite{79Leslie}. The new relation \eqref{eq:P-like}, on the contrary, is specific to the present theory of anisotropic nematic relaxation. Checking \eqref{eq:P-like} against evidence from independent sources provides therefore a significant test of our theory. \par
To do so, we have at our disposal both experimental and theoretical results, along with numerical simulations. More precisely, in what follows we analyze: (i) an early paper \cite{88hess} on MD simulation of NLCs, inspired by the model molecular theory put forward by Helfrich \cite{69helf}, and a relatively recent one \cite{07wu}, based on the Gay-Berne potential; (ii) the experimental study \cite{82kneppe} universally regarded as the standard reference for the viscosities of MBBA between $20$ and $44^\circ$C; (iii) the outcome of a study on non-equilibrium statistical mechanics initiated by Osipov and Terentjev \cite{NESM,Chrza&}, extending previous results by Kuzuu and Doi \cite{Kuzuu&Doi} (see, in particular, the recent comprehensive review by Chan and Terentjev \cite{NESM}). \par
In \cite{88hess}, Baals and Hess computed a complete set of viscosity coefficients by running a series of non-equilibrium MD simulations on a small system comprised of 128 particles, interacting through  either a Lennard-Jones ellipsoid or a soft ellipsoid (purely repulsive) potential and subjected to plane Couette flows with various shear rates and different orientations relative to the uniform nematic direction, which was kept fixed in all runs. Their results are immediately comparable with the predictions of our theory, both having been obtained for a perfectly aligned nematic fluid. The coefficients in \cite{88hess} may hence enter directly the left and right sides of equality \eqref{eq:P-like}, which happens to be satisfied remarkably well (see Table~\ref{tab:MD}). \par
All remaining data are obtained for partially oriented NCLs. Nematic viscosities depend on the degree of nematic order essentially through the scalar \emph{Maier-Saupe order parameter} $S$, ranging from 0 (isotropic state) to 1 (perfect alignment) \cite{NESM,Chrza&,Kuzuu&Doi,88Lee;95ehhe;04sovi}. To compensate for the fact that our theory does not account for partial order, we obtain a crude reconstruction of the nominal values of nematic viscosities at $S\!=\!1$ from values measured for partially ordered NLCs by replacing each $\bn\!\tp\!\bn$ term in the viscous stress \eqref{eq:TLeslie} by the corresponding second-moment tensor $S\bn\!\tp\!\bn$ \cite{88Lee;95ehhe;04sovi} and taking into account that the nematic contribution to the shear viscosity $\alpha_4$ is overshadowed by a dominant isotropic contribution \cite{82kneppe,Chrza&}. \par
In \cite{07wu}, Wu, Qian, and Zhang performed non-equilibrium MD simulations on a system of about 6,000 molecules, interacting via a Gay-Berne potential, for determining the six Leslie coefficients for each of three different shear rates. Their values, extrapolated from $S=0.75$ to $1$ through the above-described reconstruction procedure, show a striking agreement with relation \eqref{eq:P-like} (see Table~\ref{tab:MD}). \par
\begin{figure}[h]
\includegraphics[width= 0.48\textwidth]{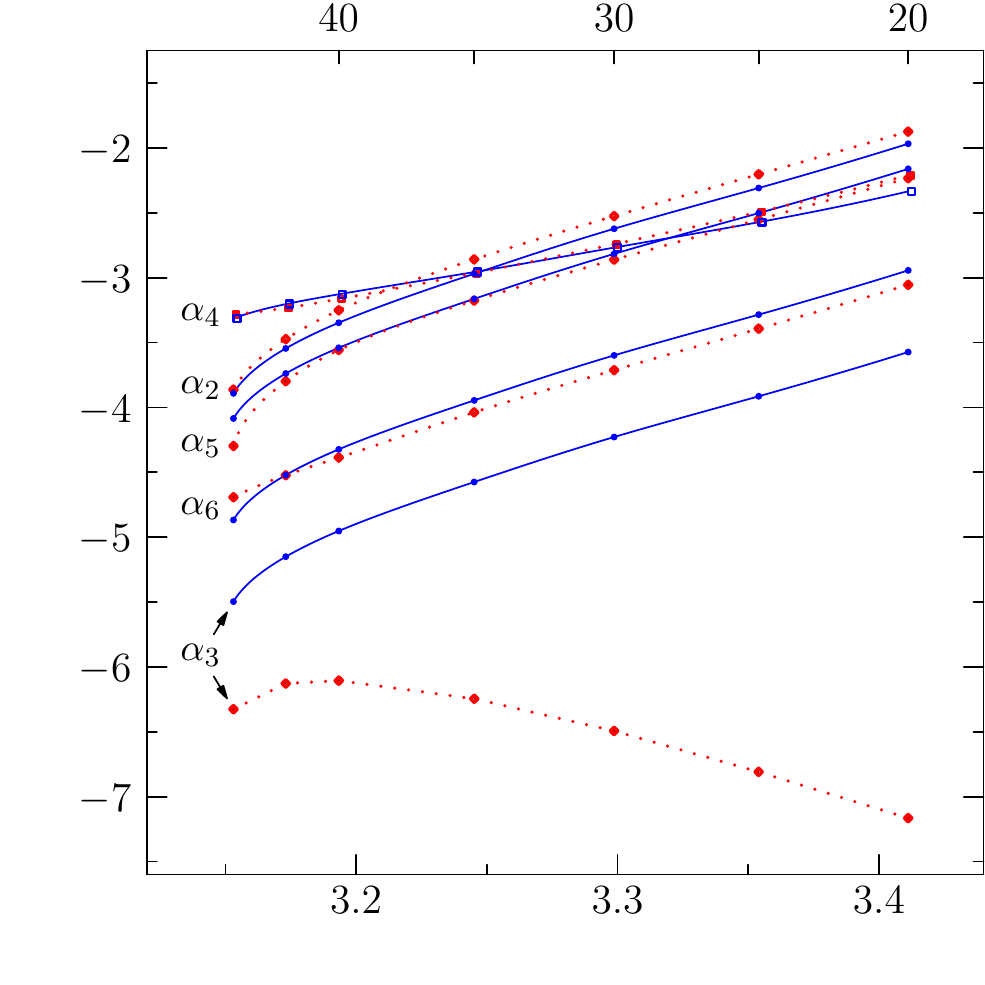}
\caption{Viscosity coefficients of nematic MBBA as a function of temperature: our $S$-rescaled values (solid lines) vs.~experimental data from \cite{82kneppe} (dotted lines). Bottom $x$-axis: inverse of absolute temperature (${\rm mK}^{-1}$); top $x$-axis: temperature ($^\circ$C); $y$-axis: logarithm of the \emph{modulus} of viscosities in Pa\,s (all of them negative, except $\alpha_4$ and $\alpha_5$).
}
\label{fig:viscosities}
\end{figure}

In \cite{82kneppe}, Kneppe et al.\ provided a complete set of Leslie coefficients for several temperature values ranging from $20$ to $44^\circ$C. Fig.\,\ref{fig:viscosities} shows that the measured values may be satisfactorily fitted with our predictions \eqref{eq:alphas}, provided we assume that the Leslie viscosities depend on the Maier-Saupe order parameter $S$ as discussed above, and $S$ itself depends on temperature as in Table~\ref{tab:MBBA}. A remarkable exception is provided by the Leslie coefficient $\alpha_3$, which deserves special attention. In fact, in \cite{82kneppe} the authors themselves raise a warning concerning this coefficient, which they derive as the difference of two nearly equal quantities, to the point that they hope for alternative measuring techniques. In particular, one of the striking peculiarities of the experimental $\alpha_3$ estimate in \cite{82kneppe} is that, at variance with all other viscosities, it appears to increase when the degree of orientation decreases. On
the contrary, our theory predicts a consistent temperature dependence for all nematic viscosities. \par
\begin{table}[t]
\centering
\begin{tabular}{|c|c|c|c|}
\hline
% Row 0
\phantom{$\Bigg|$}\!\!Source &
$\dfrac{\alpha_2}{\alpha_3}\Big/\dfrac{\alpha_4+\alpha_5}{\alpha_4+\alpha_6}$ &
$\dfrac{\alpha_2+\alpha_3}{\alpha_6-\alpha_5}$ &
$a$
\\[2.25ex] \hline%\hline
& & & \\[-2.75ex]
% Row 1
\cite{88hess} LJE &
$1.01\pm0.59$ &
$0.91\pm0.41$ &
$1.75\pm0.62$
\\[.5ex]
% Row 2
\!\!\cite{88hess}~\;SE &
1.01 $\pm$ 0.45 &
1.01 $\pm$ 0.39 &
1.78 $\pm$ 0.57
\\[.5ex]
% Row 3
\cite{07wu} 0.066 &
0.92 $\pm$ 0.24 &
1.01 $\pm$ 0.06 &
2.02 $\pm$ 0.09
\\[.5ex]
% Row 4
\cite{07wu} 0.044 &
0.98 $\pm$ 0.30 &
1.04 $\pm$ 0.08 &
1.96 $\pm$ 0.10
\\[.5ex]
% Row 5
\,\cite{07wu} 0.022\, &
1.10 $\pm$ 0.73 &
\;1.10 $\pm$ 0.20\; &
\;1.94 $\pm$ 0.21\;
\\[.75ex] \hline
\end{tabular}
\caption{LJE and SE stand respectively for Lennard-Jones ellipsoid and soft ellipsoid potentials, as used in \cite{88hess}; the shear rates used in \cite{07wu} are given in L-J reduced units. The third column, relative to Parodi's relation \eqref{eq:P}, is provided as a term of comparison. The scatter of the results from \cite{88hess} is huge, due to the small size of the molecular sample.}
\label{tab:MD}
\end{table}
\begin{table}[t]
\centering
\begin{tabular}{c|ccccccc}
$\,T \,[^\circ \text{C}\,]$\, & 20 & 25 & 30\, & 35 & 40 & 42 & 44\, \\[1mm] \hline
$S$ & \;0.92\, & \,0.66\, & \,0.48\, & \,0.34\, & \,0.23\, & \,0.19\, & \,0.14\;
\end{tabular}
\caption{Values of the Maier-Saupe order parameter $S$ identified by a best fit between experimental MBBA viscosities in \cite{82kneppe} and our theory (see text for details). Note that the temperature dependence we obtain for $S$ is consistent with a weakly first-order nematic-isotropic phase transition, with a critical temperature $T_\textrm{\tiny NI}\simeq 49\,^\circ$C.}
\label{tab:MBBA}
\end{table}
 \par
A third and final test for our theoretical predictions comes from the non-equilibrium Fokker-Planck analysis developed in \cite{NESM,Chrza&,Kuzuu&Doi}. This mean-field theory deals exclusively with the relaxational dynamics of the orientational degrees of freedom of anisotropic molecules to the exclusion of the translational relaxation associated with shear flow. Consequently, it does not provide reliable predictions for the shear viscosity coefficient $\alpha_4$, and therefore it cannot be used as a direct test for the new Parodi-like relation \eqref{eq:P-like}. Moreover, this theory delivers an explicit universal representation only for the symmetric part of the stress tensor, while the rotational viscosity $\gamma_1\!=\alpha_3-\alpha_2$  (and hence the complete set of Leslie viscosities) depends on the specific form of the assumed mean-field potential. That having been said, results in \cite{NESM,Chrza&,Kuzuu&Doi} are in good qualitative agreement with our formulas \eqref{eq:alphas}---with the obvious 
exception of $\alpha_4$. \par
Equality \eqref{eq:P-a} further allows us to establish a direct link between the aspect ratio $a$---playing a key role in the present theory of nematic relaxation but not directly observable---and quantities amenable to experimental and numerical determination. The values obtained from data in \cite{88hess,07wu} are collected in the fourth column of Table~\ref{tab:MD}.\par
\section{Discussion}
We have presented a hydrodynamic theory that accounts for both elastic and relaxational effects, based only on  material symmetry requirements. When applied to NLCs in the low-frequency regime, this theory predicts another relation beyond Parodi's linking the six Leslie viscosities, thus lowering to four the number of independent nematic viscosities. Our predictions are in remarkable quantitative accord with experimental measures on MBBA and MD simulations, and in fair qualitative agreement with earlier theoretical predictions. The basic tenet of our theory is the separation between equilibrium properties, encoded in the free energy functional, and non-equilibrium properties, encoded in the relaxation dynamics. Correspondingly, the distinction between `solid' and `liquid' rests on the activated relaxation mechanisms, and not on the underlying energetics. In fact, the strain energy functional characterizing anisotropic \mbox{(visco-\!)}elastic \emph{solids} such as nematic elastomers and the anisotropic 
potential \eqref{eq:
sheared} we use for nematic \emph{liquid} crystals are formally alike. \par
NLCs may be classified into two groups: \emph{flow-aligning} (such as MBBA and 5CB) and \emph{tumbling} (such as HBAB and 8CB) \cite{Tumble}, characterized, respectively, by a positive or negative value of the \emph{tumbling parameter},
\begin{equation}\label{eq:tumble}
\lambda=\frac{1+\alpha_3/\alpha_2}{1-\alpha_3/\alpha_2}\,.
\end{equation}
Since $a$ is intrinsically positive and reasonably greater than 1, \eqref{eq:P-like} implies $0<\alpha_3/\alpha_2<1\Leftrightarrow\lambda>0$. Therefore, our theory covers only flow-aligning NLCs. While narrowing its scope, this limitation makes it more specific. The hidden link between assumption \eqref{eq:sheared} and flow alignment surely deserves further study, as does a proper incorporation of the degree of nematic order. A separate issue we intend to address is removing the incompressibility constraint, paying due attention to the possible role of $\tau_6$ \cite{tau6}, in order to reconsider the nemato\-acoustic problem we tackled in \cite{14bdt15tur}. \par
\vspace{5.ex}
\begin{acknowledgments}
This paper is dedicated to Jerry Ericksen on the occasion of his 90th birthday. Financial support from the Italian Ministero dell'Istruzione, dell'Universit\`a e della Ricerca through the Grant No.\,200959L72B\underline{\;\;}004 ``Mathematics and Mechanics of Biological Assemblies and Soft Tissues'', is gratefully acknowledged. Independent comments from Maria-Carme~Calderer and an anonymous referee were very helpful, and they enabled us to increase the breadth and depth of our work.
\end{acknowledgments}

%

%
%:EOF

\begin{thebibliography}{99}
%
\bibitem{47fre}
J.~Frenkel, in \emph{Kinetic Theory of Liquids}, edited by R.\,H.~Fowler, P.~Kapitza, and N.\,F.~Mott (Oxford University Press, Oxford, 1947).
%
\bibitem{85hido07tra12bflrt}
R.\,M.~Hill and L.\,A.~Dissado, J.\ Phys.\ C \textbf{18},\,3829 (1985).
K.~Trachenko, Phys.\ Rev.\ B \textbf{75}, 212201 (2007).
V.\,V. Brazhkin et al, Phys.\ Rev. E \textbf{85}, 031203 (2012).
%
\bibitem{91hcg}
H.-W.~Hu, G.\,A.~Carson, and S.~Granick, Phys.\ Rev.\ Lett.\,\textbf{66}, 2758 (1991).
%
\bibitem{95gr&95kr&99ruo&12kkkk}
C.~Grammes et al, Phys.\ Rev.\ E \textbf{51}, 430 (1995).
J.\,K.~Kr\"uger et al, Phys.\ Rev.\ E \textbf{51}, 2115 (1995).
G.~Ruocco and F.~Sette, J.\ Phys.\ Condens.\ Matter \textbf{11} R259 (1999).
J.\,H.~Kim et al, J.\ Korean Phys.\ Soc.\,\textbf{61}, 862 (2012).
%
\bibitem{02diq04rasi}
A.~DiCarlo and S.~Quiligotti, Mech.\ Res.\ Commun.\,\textbf{29}, 449 (2002).
K.\,R.~Rajagopal and A.\,R.~Srinivasa, Z.\ Angew.\ Math.\ Phys.\,\textbf{55}, 861 (2004).
%
\bibitem{14bdt15tur}
P.~Biscari, A.~DiCarlo, and S.\,S.~Turzi, Soft\ Matter \textbf{10}, 8296 (2014).
S.\,S.~Turzi, Eur.\ J.\ Appl.\ Math.\,\textbf{26}, 93 (2015).
%
\bibitem{72muls}
M.\,E.~Mullen, B.~L\"uthi, and M.\,J.~Stephen, Phys.\ Rev.\ Lett.\,\textbf{28}, 799 (1972).
%
\bibitem{70lordlab}
A.\,E.~Lord Jr.\ and M.\,M.~Labes, Phys.\ Rev.\ Lett.\,\textbf{25}, 570 (1970).
%
\bibitem{60eri}
J.\,L.~Ericksen, Arch.\ Ration.\ Mech.\ Anal.\,\textbf{4}, 231 (1959).
%
\bibitem{68lesl}F.\,M.~Leslie, Arch.\ Ration.\ Mech.\ Anal.\,\textbf{28}, 265 (1968).
%
\bibitem{70paro}
O.~Parodi, J.\ Physique \textbf{31}, 581 (1970).
%
\bibitem{75mon&}
S.\,E.~Monroe, Jr. et al, J.\ Chem.\ Phys.\,\textbf{63}, 5139 (1975).
%
\bibitem{O-F}
P.-G.~de~Gennes and J.~Prost, \emph{The Physics of Liquid Crystals}, 2nd ed. (Clarendon Press, New York, 1995).
%
\bibitem{ratio}
In \cite{14bdt15tur} we called this deviation the \emph{asphericity factor}, and we denoted it by $a$.
%
\bibitem{dev}
The deviator of a double tensor $\bL$ is its traceless component: $\dev\bL\bydef\bL-\tfrac13(\tr\bL)\bI\,.$
%
\bibitem{IP}
The inner product between two double tensors $\bL,\bM$ is defined as $\bM\!\cdot\!\bL=\tr\,(\bM^{\top}\bL)$.
%
\bibitem{87povi}
P.~Podio-Guidugli and E.\,G.~Virga, Proc.\ R.\ Soc.\ London, Ser. A \textbf{411}, 85 (1987).
%
\bibitem{Col&Noll}
B.\,D.~Coleman and W.~Noll, Arch.\ Rational\ Mech.\ Anal.\,\textbf{6}, 355 (1960).
%
\bibitem{tau6}
The coefficient $\tau_6$ does affect the spherical component of the viscous stress, obliterated here by the incompressibility constraint---and hence by the $\dev$ projector in \eqref{eq:LinGradFlow}.
%
\bibitem{79Leslie}
F.\,M.~Leslie, in \emph{Advances in Liquid Crystals}, edited by G.\,H.~Brown (Academic Press, New York, 1979), Vol.\,4, p.\,34.
%
\bibitem{88hess}
D.~Baalss and S.~Hess, Z.\ Naturforsch.\ A \textbf{43}, 662 (1988).
%
\bibitem{69helf}
W.~Helfrich, J.\ Chem.\ Phys.\,\textbf{50}, 100 (1969); \textbf{53}, 2267 (1970).
%
\bibitem{07wu}
C.~Wu, T.~Qian, and P.~Zhang, Liq.\ Cryst.\,\textbf{34}, 1175 (2007).
%
\bibitem{82kneppe}
H.~Kneppe, F.~Schneider, and N.\,K.~Sharma, J.\ Chem.\ Phys.\,\textbf{77}, 3203 (1982).
%
\bibitem{NESM}
M.\,A.~Osipov and E.\,M.~Terentjev, Phys.\ Lett.\ A \textbf{134}, 301;
Z.\ Naturforsch.\ A \textbf{44}, 785 (1989).
C.\,J.~Chan and E.\,M.~Terentjev, in \emph{Modeling of Soft Matter}, edited by M.-C.\,T.~Calderer and E.\,M.~Terentjev (Springer, New York, 2005), p.\,27; J.\ Phys.\ A: Math.\ Theor. \textbf{40}, R103 (2007).
%
\bibitem{Chrza&}
A.~Chrzanowska and K.~Sokalski, Phys.\ Rev.\ E \textbf{52}, 5228 (1995).
A.~Chrzanowska, Phys.\ Rev.\ E \textbf{62}, 1431 (2000).
%
\bibitem{Kuzuu&Doi}
N.~Kuzuu and M.~Doi, J.\ Phys.\ Soc.\ Jpn. \textbf{52}, 3486 (1983); \textbf{53}, 1031 (1984).
%
\bibitem{88Lee;95ehhe;04sovi}
S.-D.~Lee, J.\ Chem.\ Phys. \textbf{88}, 5196 (1988).
H.~Ehrentraut and S.~Hess, Phys.\ Rev.\ E \textbf{51}, 2203 (1995).
A.\,M.~Sonnet, P.\,L.~Maffettone, and E.\,G.~Virga, J.\ Non-Newtonian\ Fluid\ Mech.\,\textbf{119}, 51 (2004).
%
\bibitem{Tumble}
Ch.~G\"awhiller, Phys.\ Rev.\ Lett.\,\textbf{28}, 1554 (1972).
P.\,T. Mather, D.\,S.~Pearson, and R.\,G.~Larson, Liq.\ Cryst.\,\textbf{20}, 527 (1996); \textbf{20}, 539 (1996).
J.\,F.~Fatriansyah and H.~Orihara, Phys.\ Rev.\ E \textbf{88}, 012510 (2013).
\end{thebibliography}
\end{document}